\documentclass[a4paper,11pt]{article}
\usepackage{pos,enumitem,float,pifont}
\usepackage{xcolor,soul}

\usepackage{tikz}
\usetikzlibrary{arrows.meta}
\usetikzlibrary{decorations.markings}
\tikzstyle midarrow=[postaction={decorate,decoration={markings,
      mark=at position 0.53 with {\arrow{#1}},}}]

\graphicspath{{Figures/}}
\title{Gradient Flow Renormalisation for Meson Mixing and Lifetimes}
\ShortTitle{Gradient Flow Renormalisation for Meson Mixing and Lifetimes}

\author*[a,1]{Matthew Black}
\author[b]{Robert Harlander}
\author[c,d]{Fabian Lange}
\author[e]{Antonio Rago}
\author[b,f,g]{Andrea Shindler}
\author[a]{Oliver Witzel}

\affiliation[a]{
    Center for Particle Physics Siegen, Theoretische Physik 1,
    Naturwissenschaftlich-Technische Fakult\"at,
    Universit\"at Siegen, 57068 Siegen, Germany
  }
\affiliation[b]{
    Institute for Theoretical Particle Physics and Cosmology,
    RWTH Aachen University, 52056 Aachen, Germany
  }
\affiliation[c]{
    Physik-Institut, 
    Universität Zürich, 
    Winterthurerstrasse 190, 8057 Zürich, Switzerland
  }
\affiliation[d]{
    PSI Center for Neutron and Muon Sciences,
    5232 Villigen PSI, Switzerland
  }
\affiliation[e]{
    IMADA and Quantum Theory Center,
    University of Southern Denmark, Odense, Denmark
  }
\affiliation[f]{
    Nuclear Science Division, Lawrence Berkeley National Laboratory,
    Berkeley, CA 94720, USA
    }
\affiliation[g]{
    Department of Physics, University of California, 
    Berkeley, CA 94720, USA
    }
\note{Present address: School of Physics and Astronomy, University of Edinburgh, Edinburgh EH9 3JZ, UK}

\emailAdd{matthew.black@uni-siegen.de}

\abstract{
    Fermionic gradient flow in combination with the short-flow-time expansion provides a computational method where the renormalisation of hadronic matrix elements on the lattice can be simplified to address e.g.~the issue that operators with different mass dimension can mix.

    We demonstrate our gradient flow renormalisation procedure by determining matrix elements of four-quark operators describing neutral meson mixing or meson lifetimes.
    While meson mixing calculations are well-established on the lattice and serve to validate our procedure, a lattice calculation of matrix elements for heavy meson lifetimes is still outstanding.
    Preliminary results for mesons formed of a charm and strange quark are presented.
}

\FullConference{%
  The 41st International Symposium on Lattice Field Theory (Lattice 2024)\\
  July 28th - August 3rd, 2024\\
  Liverpool, UK\\[8pt]
{\rm Preprints: SI--HEP--2024--21, P3H-24-067, ZU-TH 46/24, PSI-PR-24-19, TTK-24-37}}

\def\mev{\,\text{Me\hspace{-0.1em}V}}
\def\fm{\,\text{fm}}

\begin{document}
\maketitle

\section{Introduction}
The phenomenology of $B$ mesons is a well-studied area at many collider experiments~\cite{HFLAV:2022wzx}, with an impressive increase in precision of the measured properties of $B$ mesons~\cite{Albrecht:2024oyn}.
To fully leverage experimental results, theoretical predictions need to reach a similar precision. 
Of critical importance is to improve the precision of non-perturbative parameters.
These can e.g.~be calculated in terms of hadronic matrix elements using lattice QCD.
Matrix elements of four-quark dimension-six operators are important in further constraining the behaviour of neutral meson mixing and also predicting the lifetime of a $B$ meson via the heavy quark expansion (HQE).

The calculation of short-distance $\Delta Q=2$ (for generic heavy quark $Q$) operators governing neutral meson mixing is well established on the lattice for both charm~\cite{Carrasco:2014uya,Carrasco:2015pra,Bazavov:2017weg} and bottom~\cite{Carrasco:2013zta,Aoki:2014nga,Gamiz:2009ku,Bazavov:2016nty,Dowdall:2019bea,Boyle:2021kqn} sectors.
The $\Delta Q=0$ matrix elements relevant for lifetime predictions have received less attention on the lattice.
There exist early quenched~\cite{DiPierro:1998ty,DiPierro:1999tb} and preliminary unquenched~\cite{Becirevic:2001fy} results from the turn of the millennium, however nothing else for many years until more recently~\cite{Lin:2022fun,Black:2023vju,Black:Thesis24}.
While some of the calculation follows similarly to the case of the $\Delta Q=2$ operators, the $\Delta Q=0$ operators additionally require disconnected and `eye' diagrams where the statistical noise is much larger.
Furthermore, mixing with operators of lower mass dimension arises under renormalisation.

In the following, we study the gradient flow (GF)~\cite{Narayanan:2006rf,Luscher:2010iy,Luscher:2013cpa} as a non-perturbative tool to simplify the renormalisation procedure in lattice calculations.
We match the results at finite flow time, $\tau$, to the $\overline{\rm MS}$ scheme using the short-flow-time expansion (SFTX)~\cite{Luscher:2011bx,Suzuki:2013gza,Luscher:2013vga} where the matching coefficients $\zeta^{-1}$ are calculated perturbatively~\cite{Endo:2015iea,Hieda:2016lly,Harlander:2022tgk,Borgulat:2023xml}.
We first test our method using the $\Delta Q=2$ matrix elements where findings can be validated against the literature, and then show first results towards a lattice calculation of the $\Delta Q=0$ matrix elements.

\section{Lattice calculation}
\vspace{-10pt}
We use six RBC/UKQCD $2{+}1$-flavour domain-wall fermion (DWF) and Iwasaki gauge ensembles with three lattice spacings $a\sim 0.11$, $0.08$, $0.07\fm$ and $267\mev\leq m_\pi\leq433\mev$ as determined by RBC/UKQCD~\cite{Blum:2014tka,Boyle:2017jwu,Boyle:2018knm}. 
These ensembles are listed in Table~\ref{tab:ensembles}. 
Light and strange quarks are simulated with the Shamir kernel of the DWF action \cite{Kaplan:1992bt,Shamir:1993zy,Furman:1994ky,Blum:1996jf} with $M_5=1.8$. 
\begin{table}[th]
    \centering
    \begin{tabular}{lcccccccccccc}
        \hline\hline & $L$ & $T$ & $a^{-1}\!/\!$GeV & $am_l^\text{sea}$ & $am_s^\text{sea}$ 
                     & $am_c^{\rm val}$ & $m_\pi/\!$MeV & srcs$\times {N}_{\text{conf}}$ & $\sigma$ & $N_\sigma$ \\\hline\hline
        C1  & $24$ & $64$ & $1.7848$ & $0.005$    & $0.040$ & $0.64$ & $340$ & $32\times101$ & 4.5 & 400 \\
        C2  & $24$ & $64$ & $1.7848$ & $0.010$    & $0.040$ & $0.64$ & $433$ & $32\times101$ & 4.5 & 400 \\[1.2ex]
        M1  & $32$ & $64$ & $2.3833$ & $0.004$    & $0.030$ & $0.45$ & $302$ & $32\times\phantom{0}79$ & 6.5 & 400 \\
        M2  & $32$ & $64$ & $2.3833$ & $0.006$    & $0.030$ & $0.45$ & $362$ & $32\times\phantom{0}89$ & 6.5 & 100 \\
        M3  & $32$ & $64$ & $2.3833$ & $0.008$    & $0.030$ & $0.45$ & $411$ & $32\times\phantom{0}68$ & 6.5 & 100 \\[1.2ex]
        F1S & $48$ & $96$ & $2.785$  & $0.002144$ & $0.02144$ & $0.37$ & $267$ & $24\times\phantom{0}98$ & & 
        \\\hline\hline 
    \end{tabular}
  \caption{RBC/UKQCD ensembles used in the discussed simulations~\cite{Allton:2008pn,Aoki:2010dy,Blum:2014tka,Boyle:2017jwu}.
    $am_l^\text{sea}$ and $am_s^\text{sea}$ are the light and strange sea quark masses and $m_\pi$ is the unitary pion mass. 
    $am_s^{\rm val}$ are the valence strange quark masses, set to the physical mass.}
  \label{tab:ensembles}
\end{table}

Heavy quarks are simulated using stout-smeared gauge fields \cite{Morningstar:2003gk}~and the M\"obius kernel of the DWF action \cite{Brower:2012vk} with parameters $b=1.5$ and $c=0.5$, where the mass has been tuned to the physical charm mass on each ensemble through the $D_s$ pseudoscalar meson \cite{ParticleDataGroup:2022pth}.
Using a similar setup as Ref.~\cite{Boyle:2018knm}, all propagators are generated with Z2-noise wall sources to which we apply Gaussian smearing for the strange quarks on the C and M ensembles.
The number of sources and smearing parameters are listed in Table~\ref{tab:ensembles}.
Measurements were performed using \texttt{Grid}~\cite{GRID,Grid16} and \texttt{Hadrons}~\cite{Hadrons22}.

\begin{figure}[th]
    \centering
    \includegraphics[width=1.0\textwidth]{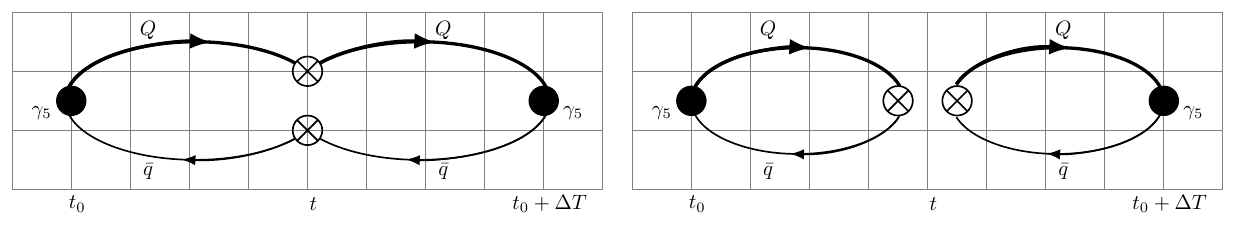}
    \caption{\label{fig:quarklines} Quark line diagrams for the three-point correlation functions with $\Delta Q=0$ four-quark operators inserted at time $t$ between two sources at $t_0$ and $t_0+\Delta T$ for ${\cal O}_1$ (left) and $\tau_1$ (right). The two $\Delta Q=2$ diagrams for $Q_1$ follow similarly with charge inversion on one side of the operator insertion.}
\end{figure}

The operators and their bag parameters considered are
\begin{align}
    \Delta Q=2:~ Q_1 &= (\bar{q}\gamma_\mu(1-\gamma_5)Q)(\bar{q}\gamma_\mu(1-\gamma_5)Q), &
    B_1^{\Delta Q=2} &= \frac{\langle P|Q_1|P\rangle}{\eta m^2f^2}, \\
    \Delta Q=0:~{\cal O}_1 &= (\bar{Q}\gamma_\mu(1-\gamma_5)q)(\bar{q}\gamma_\mu(1-\gamma_5)Q), &
    B_1^{\Delta Q=0} &= \frac{\langle P|{\cal O}_1|P\rangle}{\eta m^2f^2}, \\
    T_1 &= (\bar{Q}\gamma_\mu(1-\gamma_5)T^aq)(\bar{q}\gamma_\mu(1-\gamma_5)T^aQ), &
    \epsilon_1 &= \frac{\langle P|T_1|P\rangle}{\eta m^2f^2}, 
\end{align}
where $q$ indicates a light quark flavour and $Q$ a heavy quark flavour. 
$P$ is the pseudoscalar heavy-light meson formed from these with mass $m$ and decay constant $f$. 
$\eta=\frac83$ for $Q_1$ and $1$ for ${\cal O}_1$ and $T_1$.
For convenience in the lattice simulation, we choose a basis without the colour generator and relate back to the original operator $T_1$ in the parity-even projection, i.e.
\begin{align}
    T_1^+ &= - \frac12 \tau_1^+ - \frac{1}{2N_c}{\cal O}_1^+, \\
    \tau_1 &= (\bar{Q}\gamma_\mu(1-\gamma_5)Q)(\bar{q}\gamma_\mu(1-\gamma_5)q).
\end{align}
The $\epsilon_1$ bag parameter is therefore not immediately available and is extracted via
\begin{equation}
    \epsilon_1 = -\frac12 \epsilon_1^{\langle\tau_1\rangle} - \frac{1}{2N_c}B_1^{\Delta Q=0},
\end{equation}
where the first term is the bag parameter of the $\tau_1$ operator in the standard definition.
The bag parameters of $Q_1$, ${\cal O}_1$, and $\tau_1$ are extracted from commonly-used ratios of three- and two-point functions; see e.g.~Ref.~\cite{Black:2023vju}.
The quark line diagrams for ${\cal O}_1$ and $\tau_1$ are shown in figure~\ref{fig:quarklines}.

The $\Delta Q=0$ operators have additional `eye' diagram topologies which are computationally more challenging; see e.g.~Ref.~\cite{RBC:2022ddw}.
Since these are predicted to be small from sum rules~\cite{King:2021jsq}, we do not yet consider them in this pilot study.
Further, while the renormalisation at finite flow time is multiplicative, the matching to $\overline{\rm MS}$ for the full $\Delta Q=0$ operators involves mixing with lower-dimensional operators. 
At this stage we do not account for operator mixing and the matching uses a simplification valid for the difference of $\Delta Q=0$ operators with different spectator quarks.

\section{Results}
\subsection{Gradient Flow Evolution of Bag Parameters}
\begin{figure}[t]
    \centering
    \includegraphics[width=0.4\textwidth]{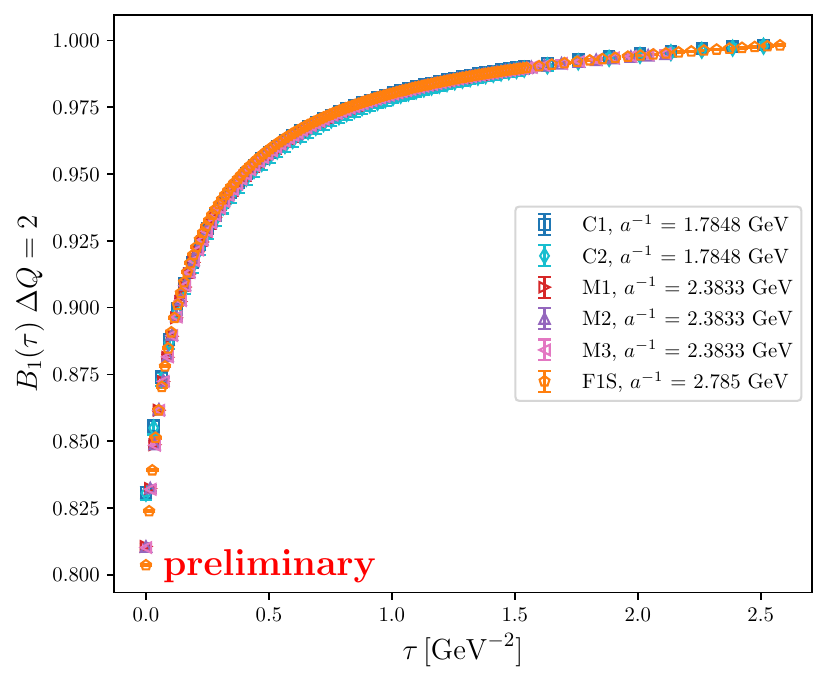}
    \caption{\label{fig:D20O1_physGF} Flow time evolution of the bag parameter $B_1$ for $\Delta Q=2$ in physical units $\tau\,[{\rm GeV}^{-2}]$.}
\end{figure}

\begin{figure}[th]
    \centering
    \includegraphics[width=0.4\textwidth]{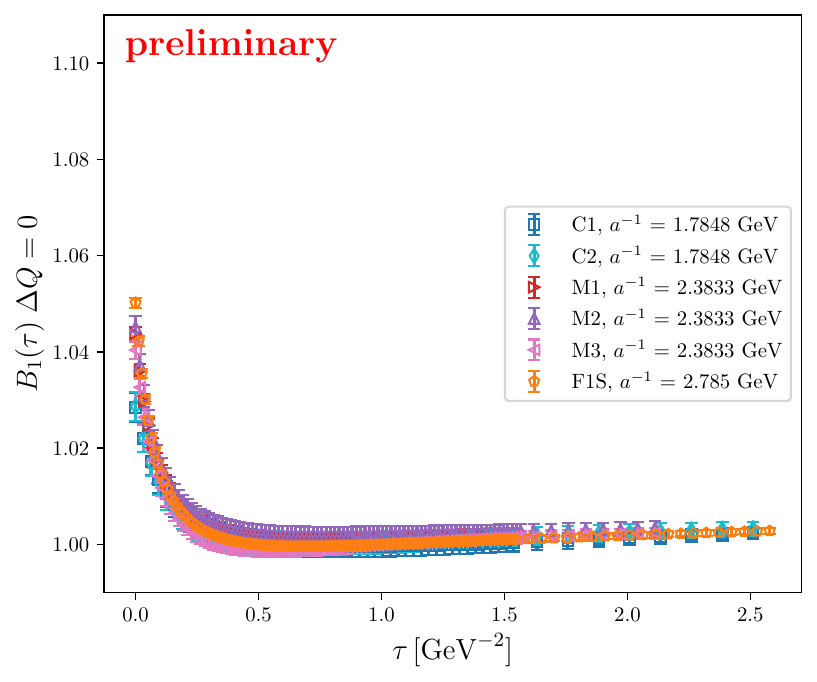}~~~~~
    \includegraphics[width=0.4\textwidth]{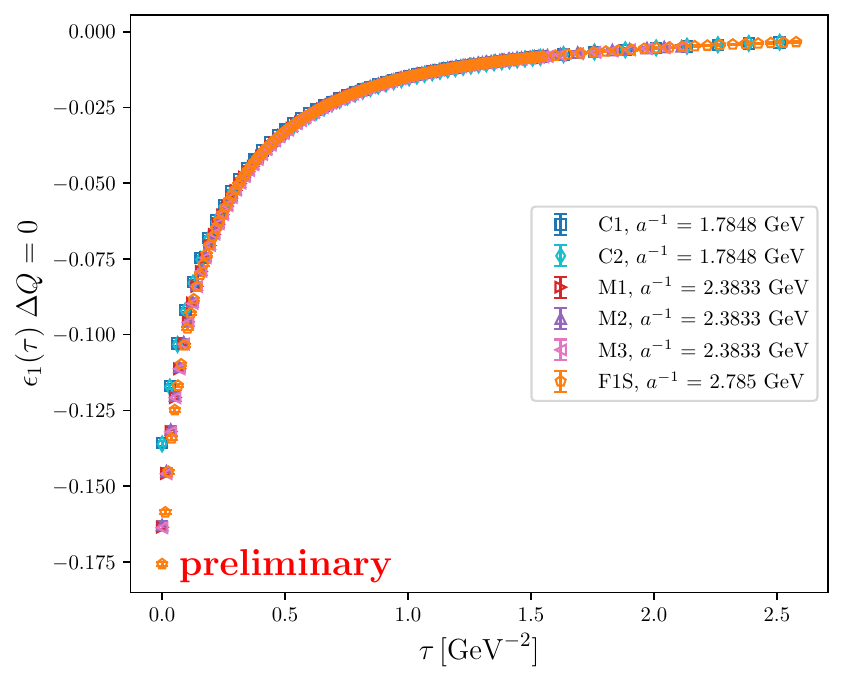}
    \caption{\label{fig:D0T1_physGF} Flow time evolution of the $\Delta Q=0$ bag parameters $B_1$ (left) and $\epsilon_1$ (right) in physical units $\tau\,[{\rm GeV}^{-2}]$.}
\end{figure}

Correlated fits are performed to extract bag parameters for each discrete flow time simulated. 
The evolution of the $B_1$ bag parameter for $\Delta Q=2$ along the flow time is shown in figure~\ref{fig:D20O1_physGF}, while the evolutions of the $\Delta Q=0$ bag parameters $B_1$ and $\epsilon_1$ are shown in figure~\ref{fig:D0T1_physGF}.
We see that the shape of evolution for $\epsilon_1$ appears almost identical to that of $B_1^{\Delta Q=2}$, while $B_1^{\Delta Q=0}$ offers a different form.
For all three quantities, we observe the following: 
\begin{enumerate}
    \item Data corresponding to different ensembles at the same lattice spacing but different light sea quark masses overlap. Hence we infer that sea quark effects are negligible;
    \item There is also overlap of data determined on ensembles with different lattice spacings, which leads to the expectation that continuum limits will be very mild. 
\end{enumerate}

\subsection{Continuum Limit}
After linearly interpolating the values of the C and M ensembles from their own set of discrete flow times to values used on the F1S ensemble, we perform the continuum limit for each operator at each discrete flow time step using a linear ansatz in $a^2$.
\begin{figure}[t]
    \centering
    \includegraphics[width=0.4\textwidth]{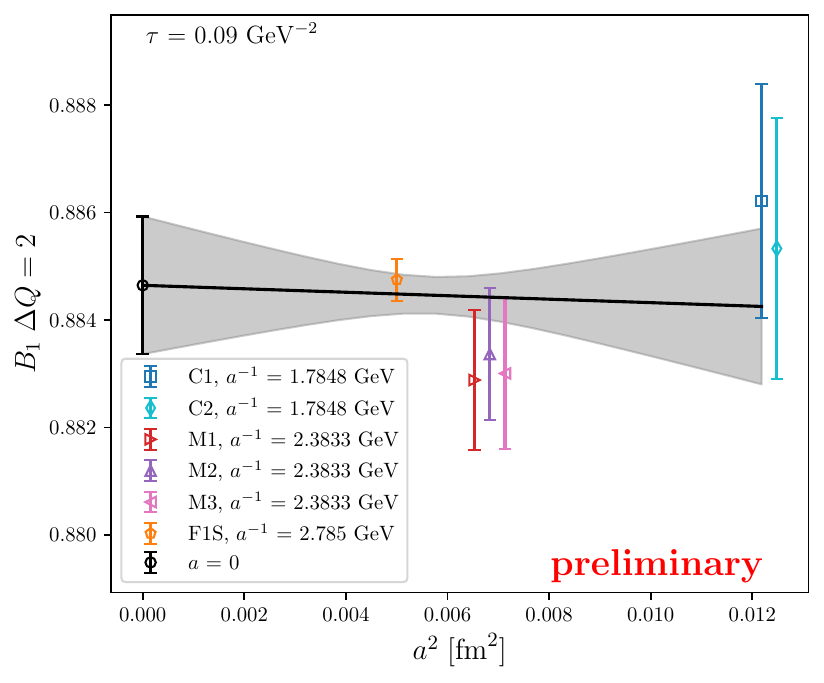}~~~~~
    \includegraphics[width=0.4\textwidth]{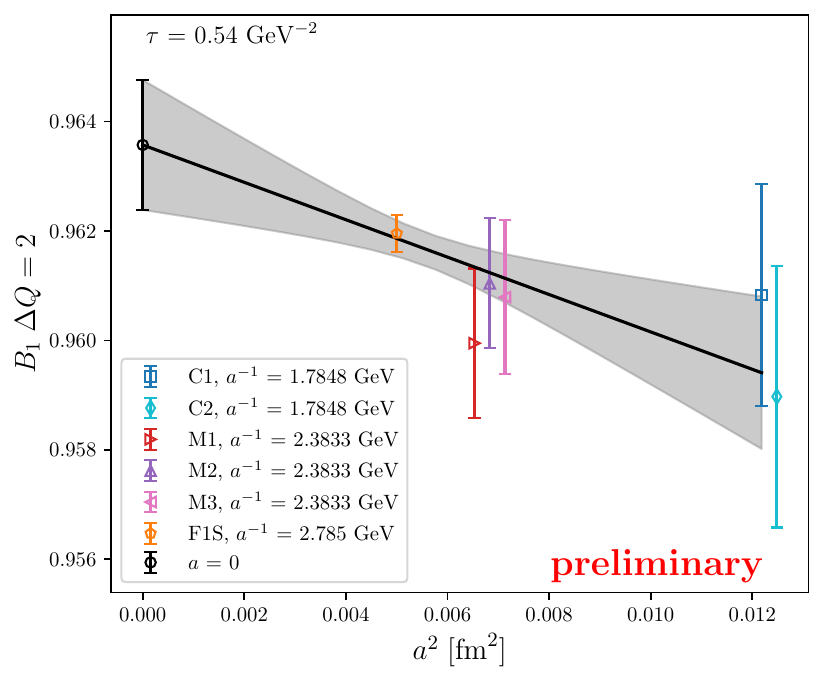}
    \caption{\label{fig:D2O1_a2fit} Examples of continuum limit extrapolations for the $B_1$ bag parameter for $\Delta Q=2$ at flow times $\tau=0.09\,$GeV$^{-2}$ (left) and $\tau=0.54\,$GeV$^{-2}$ (right). For visibility, the M1 ensemble (red) is plotted with a slight offset to left, and the C2 (cyan) and M3 (pink) ensembles to the right.}
\end{figure}
Examples of the continuum limit extrapolations at two flow times are shown for the $B_1^{\Delta Q=2}$ bag parameter in figure~\ref{fig:D2O1_a2fit}, where we present one of the `flattest' extrapolations on the left and one of the steeper extrapolations on the right. 
As suggested by the flow time evolution plots above, the continuum limits for the $\epsilon_1$ bag parameter look very similar to these.

\begin{figure}[t]
    \centering
    \includegraphics[width=0.4\textwidth]{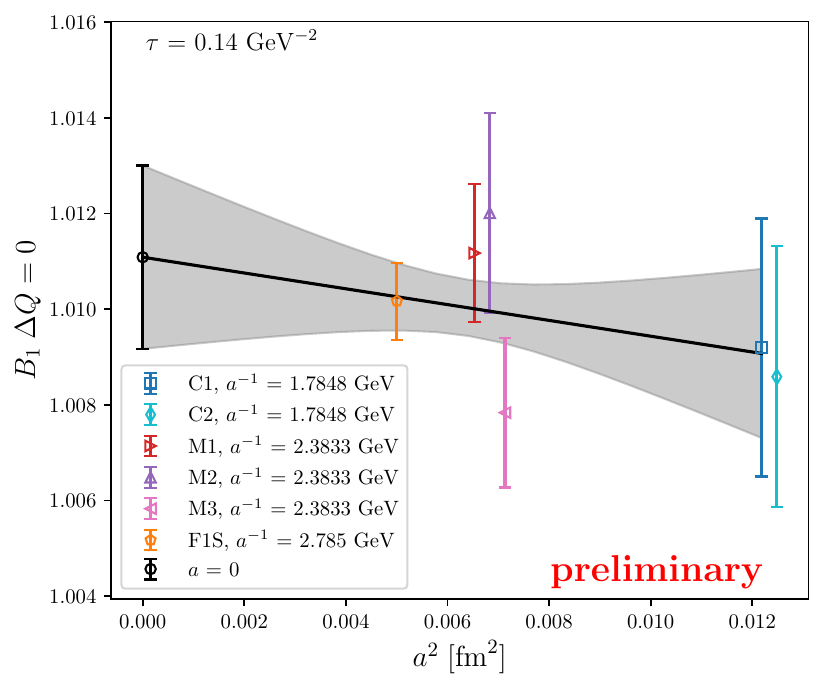}~~~~~
    \includegraphics[width=0.4\textwidth]{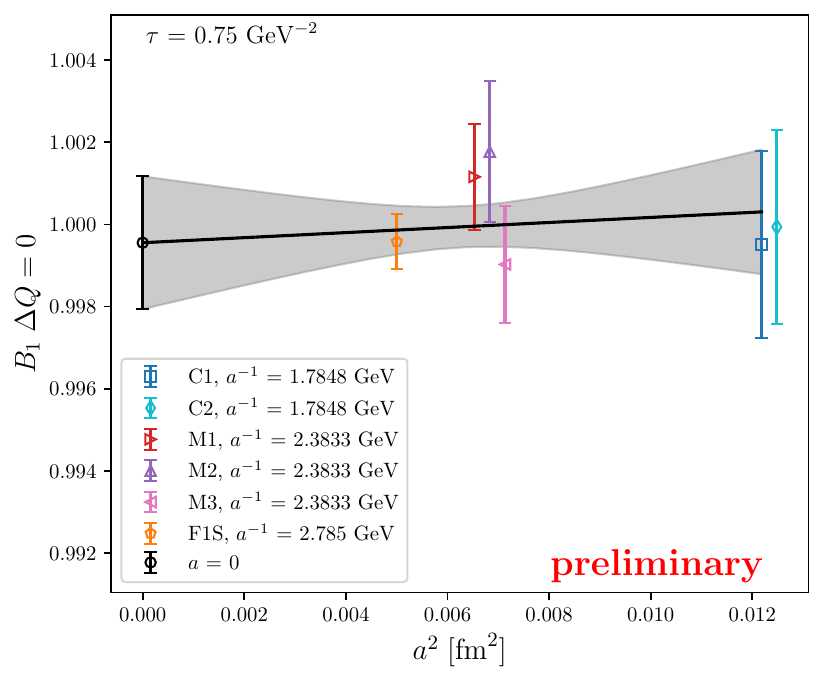}
    \caption{\label{fig:D0T1_a2fit} Examples of continuum limit extrapolations for the $B_1$ bag parameter for $\Delta Q=0$ at flow times $\tau=0.14\,$GeV$^{-2}$ (left) and $\tau=0.75\,$GeV$^{-2}$ (right). For visibility, the M1 ensemble (red) is plotted with a slight offset to left, and the C2 (cyan) and M3 (pink) ensembles to the right.}
\end{figure}

Examples of continuum limits for $B_1^{\Delta Q=0}$ are shown in figure~\ref{fig:D0T1_a2fit}.
While the evolution in the flow time has a different functional form to the other parameters, we again find relatively mild continuum limits, where the plots shown represent some of the `steepest' slopes found in the data.

\subsection{Matching to $\overline{\rm MS}$}
Having performed the continuum limit for each flow time, the final step is to match these GF-renormalised results to the $\overline{\rm MS}$ scheme.
This is done by combining flowed operators with a perturbative matching matrix $\zeta^{-1}$ which leads to the $\overline{\rm MS}$ result in the limit of $\tau\to0$.
The matching coefficients can be constructed from Refs.~\cite{Endo:2015iea,Hieda:2016lly,Harlander:2022tgk,Borgulat:2023xml} through NNLO (in QCD).
For $\Delta Q=0$, the ${\cal O}_1$ and $T_1$ operators mix such that both flowed operators contribute to the matched results.
The $\tau\to0$ limit will be taken assuming a linear extrapolation within an appropriate window in flow time. 
The flow time must be chosen sufficiently large that the data are not affected by large cut-off effects but also not too large such that the SFTX is still valid and higher-power effects remain negligible.
Within the flow time window, currently an uncorrelated linear fit is performed and then extrapolated to $\tau=0$.
We take the difference between the fit of the central values and the fit to central values $\pm1\sigma$ uncertainties to obtain the error on the extrapolated values.
The renormalisation scale is fixed to be $\mu=3\,$GeV.

\begin{figure}[t]
    \centering
    \includegraphics[width=0.7\textwidth]{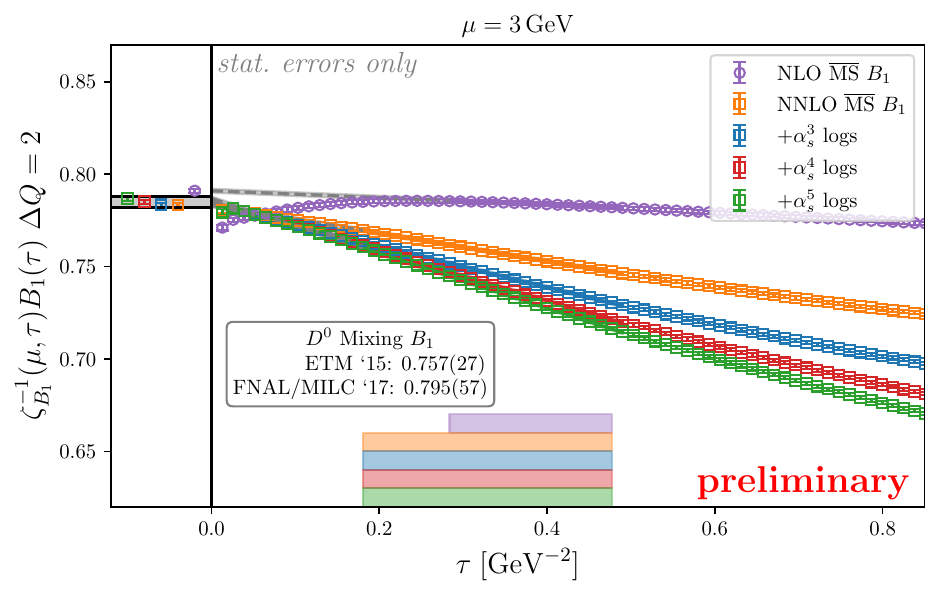}
    \caption{\label{fig:D2O1_MSb} Flow-time dependence of the combination $\zeta^{-1}_{B_1}(\mu,\tau)B_1(\tau)$ with different perturbative orders: NLO (purple), NNLO (orange), $+\alpha_s^3$ logs (blue), $+\alpha_s^4$ logs (red), $+\alpha_s^5$ logs (green). Error bars represent statistical uncertainties only. The gray bands leading from each coloured data set represent the $\tau\to0$ extrapolations taken from uncorrelated linear fits; the results at $\tau=0$ are then shown in the left panel.
    The coloured bands at the edges of the plots indicate the fit range of $\tau$ used for the short-flow-time expansion at each perturbative treatment.}
\end{figure}

First we consider the $B_1^{\Delta Q=2}$ bag parameter combined with its perturbative matching in figure~\ref{fig:D2O1_MSb}. 
We take the perturbative matching coefficient $\zeta_{B_1}^{-1}$ at both NLO and NNLO to study the systematic effects of the truncation of perturbation theory.
Beyond NNLO, we also include higher-order logarithmic terms of the form $\alpha_s^n\ln^{n-k}(t),\,k=0,1,2$, which can be derived from renormalisation group considerations.
These could be resummed to all orders in $n$, which we defer to future work. 
At NLO we choose the flow time window $0.28\,{\rm GeV}^{-2}\leq\tau\leq0.49\,{\rm GeV}^{-2}$ and at NNLO (as well as with higher logarithms) we choose $0.18\,{\rm GeV}^{-2}\leq\tau\leq0.49\,{\rm GeV}^{-2}$. 
As can be seen in the figure, the main impact of including higher-order effects in perturbation theory appears to be to extend the region of applicability for the flow time window towards smaller $\tau$.

Since we are simulating a `charm-strange' meson, the $\Delta Q=2$ bag parameters do not have proper physical meaning. 
These can however be considered as a proxy to the short-distance effects of $D^0$ meson mixing, where we assume spectator effects to be small.
In the literature, the short-distance matrix elements for $D^0$ mixing have been calculated on the lattice by FNAL/MILC at $N_f=2+1$ and ETMC at $N_f=2+1+1$, with $\mu=3\,$GeV.
ETMC finds a value of $B_1^{\overline{\rm MS}}=0.757(27)$~\cite{Carrasco:2015pra}.
FNAL/MILC quotes a value for $\langle{\cal O}_1\rangle^{\overline{\rm MS}}$ which, using PDG~\cite{ParticleDataGroup:2022pth}, translates to $B_1^{\overline{\rm MS}}=0.795(57)$~\cite{Bazavov:2017weg}.
There is also a prediction from HQET sum rules, yielding $B_1^{\overline{\rm MS}}=0.654^{+0.060}_{-0.052}$~\cite{Kirk:2017juj}.
The preliminary results shown here in figure~\ref{fig:D2O1_MSb} lie between the two lattice values and slightly above that from HQET sum rules. 
To incorporate both the statistical uncertainty of the data and the systematic error of truncation in perturbation theory, we take the full spread of the different perturbative treatments at NNLO including higher logs as the range of our final value and choose to symmetrise the errors, yielding
\begin{equation}
    B_1^{\Delta Q=2,\overline{\rm MS}}(3\,{\rm GeV}) = 0.785(3).
\end{equation}
While further systematic uncertainties are still to be included, this agreement with existing values is promising.

\begin{figure}[t]
    \centering
    \includegraphics[width=0.47\textwidth]{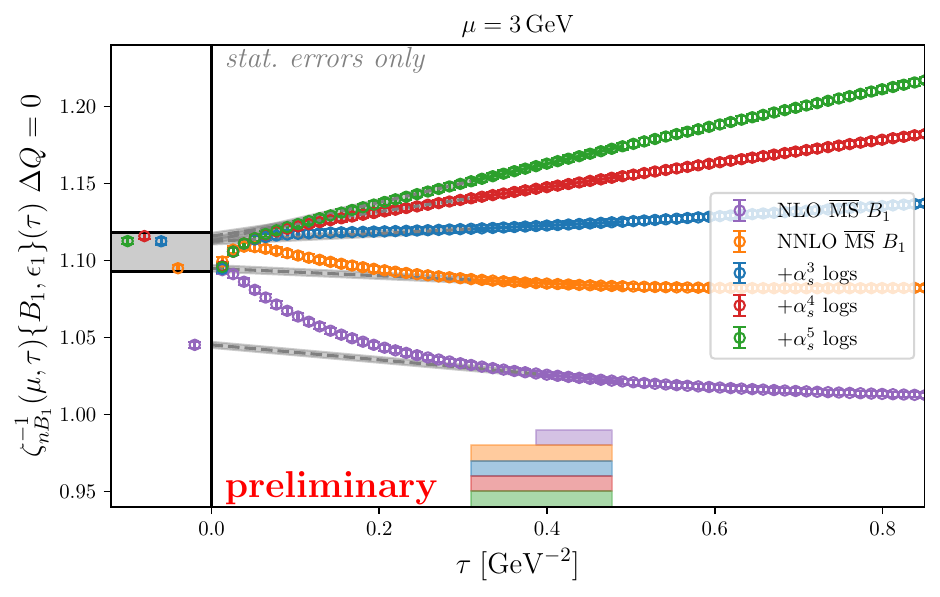}
    \includegraphics[width=0.47\textwidth]{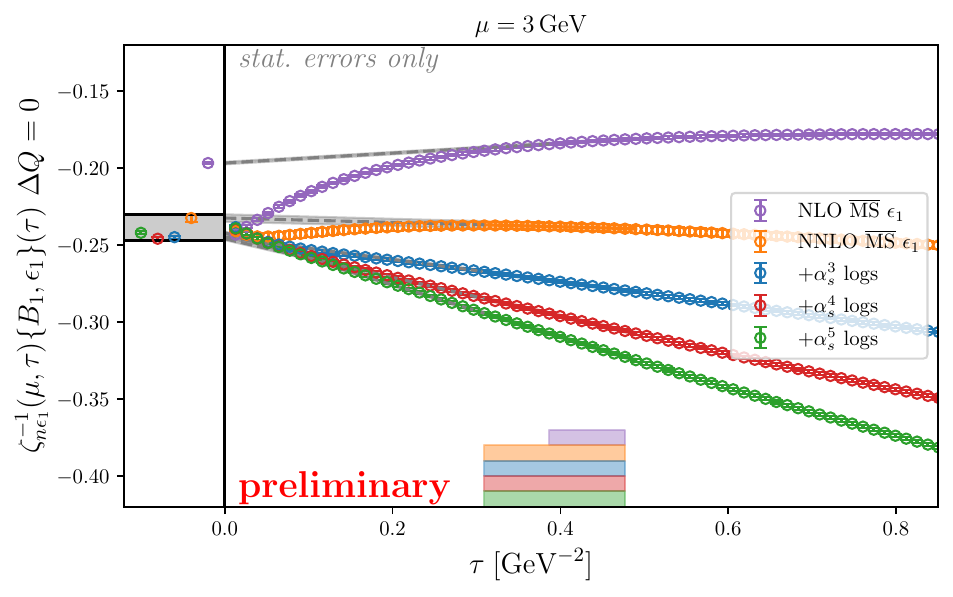}
    \caption{\label{fig:D0_MSb} 
    Flow-time dependence of the linear combinations $\zeta^{-1}_{nm}(\mu,\tau)\{B_1,\epsilon_1\}(\tau)$ for $B_1^{\Delta Q=0,\overline{\rm MS}}$ (left) and $\epsilon_1^{\overline{\rm MS}}$ (right) using different perturbative orders: NLO (purple), NNLO (orange), $+\alpha_s^3$ logs (blue), $+\alpha_s^4$ logs (red), $+\alpha_s^5$ logs (green). Error bars represent statistical uncertainties only. The gray bands leading from each coloured data set represent the $\tau\to0$ extrapolations taken from uncorrelated linear fits; the results at $\tau=0$ are then shown in the left panel.
    The coloured bands at the edges of the plots indicate the fit range of $\tau$ used for the short-flow-time expansion at each perturbative treatment.}
\end{figure}

Next we consider the $B_1^{\Delta Q=0}$ and $\epsilon_1$ bag parameters.
Although the continuum limit at fixed flow time avoids mixing with lower-dimensional operators, power divergences may still emerge in the SFTX; see e.g.~Ref.~\cite{Kim:2021qae}.
We plan to address this issue in the future. 
In the meantime, we use the perturbative matching calculated for the difference of $\Delta Q=0$ operators with different spectator quarks, where the troublesome terms cancel.
The combinations of the matching matrices and the continuum-limit lattice data are shown in figure~\ref{fig:D0_MSb} for $B_1^{\Delta Q=0,\overline{\rm MS}}$ and $\epsilon_1^{\overline{\rm MS}}$.
For both bag parameters, the flow time windows $0.39\,{\rm GeV}^{-2}\leq\tau\leq0.48\,{\rm GeV}^{-2}$ at NLO and $0.31\,{\rm GeV}^{-2}\leq\tau\leq0.48\,{\rm GeV}^{-2}$ at NNLO are chosen.
To date no lattice QCD determination with a full error budget exists, only sum rules computations in HQET~\cite{Kirk:2017juj}.
We thus decided to compare with the HQET sum rules results for lifetime differences matched to full QCD, which give $B_1^{\Delta Q=0}=0.902^{+0.077}_{-0.051}$ and $\epsilon_1=-0.132^{+0.041}_{-0.046}$~\cite{Kirk:2017juj}.
At the preliminary stages of our calculation, again incorporating the full spread of the different perturbative treatments at NNLO into the final values, we find
\begin{align}
    B_1^{\Delta Q=0,\overline{\rm MS}}(3\,{\rm GeV}) = 1.105(13), ~~~~~~\text{and}~~~~~~~~~
    \epsilon_1^{\overline{\rm MS}}(3\,{\rm GeV}) = -0.239(8),
\end{align}
and we observe that both $B_1^{\Delta Q=0}$ and $\epsilon_1$ lie relatively close to the predictions for lifetime differences based on sum rules.

\section{Summary}
For quantities deduced using the heavy quark expansion, $\Delta B=0$ four-quark matrix elements play an important role in improving precision and accuracy.
However, these are not yet determined using lattice QCD.
One of the difficulties impeding their calculation is mixing with lower-dimension operators. 
We have here demonstrated the use of the gradient flow to non-perturbatively calculated renormalised matrix elements of four-quark operators.
We convert the results at finite flow time to the $\overline{\rm MS}$ scheme by performing a perturbative matching using the SFTX.
Preliminary results have been obtained for $D_s$ mesons.
While we consider both $\Delta Q=2$ and $\Delta Q=0$ operators, only $\Delta Q=0$ have a direct physical meaning. 
A full systematic error analysis is yet to be undertaken. 

Calculating $\Delta Q=2$ bag parameters is well-established and provides a test case for our method where findings can be compared to results for short-distance $D^0$ mixing, assuming spectator effects to be negligible.
We find good agreement between our results and literature values.
Furthermore, we have made first steps towards predictions for the $\Delta Q=0$ bag parameters.
While additional operators and diagrams are still required in both the lattice calculation and perturbative matching, we observe that our preliminary results have the expected order of magnitudes.

\acknowledgments 
\vspace{-10pt}
\noindent
These computations used resources provided by the OMNI cluster at the University of Siegen, the HAWK cluster at the High-Performance Computing Center Stuttgart, and LUMI-G at the CSC data center Finland (DeiC National HPC g.a.~DEIC-SDU-L5-13 and DEIC-SDU-N5-2024053).
We thank the RBC/UKQCD collaboration for generating and making their gauge ensembles publicly available.
M.B., R.H., O.W. received support from the Deutsche Forschungsgemeinschaft (DFG, German Research Foundation) through grant 396021762 - TRR 257 ``Particle Physics Phenomenology after the Higgs Discovery''.
A.S.~received support from the DFG through grant 513989149, the National Science Foundation grant PHY-2209185 and the DOE Topical Collaboration ``Nuclear Theory for New Physics'' award No. DE-SC0023663.
The work of F.L.~was supported by the Swiss National Science Foundation (SNSF) under contract \href{https://data.snf.ch/grants/grant/211209}{TMSGI2\_211209}.
Special thanks is given to Felix Erben, Ryan Hill, and J.~Tobias Tsang for assistance in setting up the simulation code.

\section*{Comment}
\vspace{-10pt}
\noindent
After submitting the original version of this proceedings to arXiv, we discovered a mistake in our analysis for the lifetimes which has been corrected in version 2 and the resulting values for $B_1^{\Delta Q=0}$ and $\epsilon_1$ have been updated.
\vfill

\pagebreak
\bibliographystyle{JHEP-jmf-arxiv}
\setlength{\bibsep}{2pt plus 0.3ex}
\bibliography{B_meson}

\end{document}